\documentclass[journal,twoside]{IEEEtran}
\IEEEoverridecommandlockouts
\usepackage{cite}
\usepackage{amsmath,amssymb,amsfonts}
\usepackage{algorithmic}
\usepackage{graphicx}
\usepackage{textcomp}
\usepackage{xcolor}
\usepackage{booktabs}
\usepackage{fancyhdr} 

\fancypagestyle{ieee}{ 
    \fancyhf{} 
    \fancyfoot[C]{
    \footnotesize This work has been submitted to the IEEE for possible publication. \\ Copyright may be transferred without notice, after which this version may no longer be accessible.}

}

\def\BibTeX{{\rm B\kern-.05em{\sc i\kern-.025em b}\kern-.08em
    T\kern-.1667em\lower.7ex\hbox{E}\kern-.125emX}}
\begin{document}

\bstctlcite{IEEEexample:BSTcontrol}

\title{3D-TrIM: A Memory-Efficient Spatial Computing Architecture for Convolution Workloads

\thanks{This work was supported by the Engineering and Physical Sciences Research Council (EPSRC) Programme Grant "Functional Oxide Reconfigurable Technologies" (FORTE) under Grant EP/R024642/2, by the RAEng Chair in Emerging Technologies under Grant CiET1819/2/93, and by the EPSRC AI for Productive Research \& Innovation in eLectronics (APRIL) Hub under Grant EP/Y029763/1.}
\thanks{The authors are with the Centre for 
Electronics Frontiers, Institute for Integrated Micro and Nano Systems,
School of Engineering, The University of Edinburgh, EH9 3BF, Edinburgh, 
United Kingdom. (e-mails: csestito@ed.ac.uk; a.j.abdelmaksoud@ed.ac.uk; shady.agwa@ed.ac.uk; 
t.prodromakis@ed.ac.uk).}

}

\author{Cristian~Sestito, 
        Ahmed~J.~Abdelmaksoud,
        Shady~Agwa,
        and~Themis~Prodromakis}

\maketitle
\thispagestyle{ieee} 

\begin{abstract}

The Von Neumann bottleneck, which relates to the energy cost of moving data from memory to on-chip core and vice versa, is a serious challenge in state-of-the-art AI architectures, like Convolutional Neural Networks' (CNNs) accelerators. 
Systolic arrays exploit distributed processing elements that exchange data with each other, thus mitigating the memory cost. However, when involved in convolutions, data redundancy must be carefully managed to avoid significant memory access overhead. To overcome this problem, TrIM has been recently proposed. It features a systolic array based on an innovative dataflow, where input feature map (ifmap) activations are locally reused through a triangular movement. However, ifmaps still suffer from memory accesses overhead.
This work proposes 3D-TrIM, an upgraded version of TrIM that addresses the memory access overhead through few extra shadow registers. In addition, due to a change in the architectural orientation, the local shift register buffers are now shared between different slices, thus improving area and energy efficiency. An architecture of 576 processing elements is implemented on commercial 22 nm technology and achieves an area efficiency of 4.47 TOPS/mm$^2$ and an energy efficiency of 4.54 TOPS/W. Finally, 3D-TrIM outperforms TrIM by up to $3.37\times$ in terms of operations per memory access considering CNN topologies like VGG-16 and AlexNet.

\end{abstract}


\section{Introduction}

\IEEEPARstart{N}{owadays}, hardware architectures for Artificial Intelligence (AI) require vast amount of data to deliver tasks effectively. Convolutional Neural Networks (CNNs) manage multi-dimensional input feature maps (ifmaps) and filters to retrieve meaningful patterns from inputs\cite{Sze_17,Li_22}. Ifmaps and filters are initially stored in an external memory and then loaded on-chip to the processing core. However, the data transfer between memory and core requires noticeable energy: one memory access dissipates from two to three orders of magnitude higher energy than a single computation\cite{Horowitz_14}. Therefore, hardware architectures that reduce the number of memory accesses are desirable\cite{Capra_20}.

Systolic Arrays (SAs) are spatial architectures that maximize data utilization on-chip, thus accessing memory less frequently\cite{Xu_23}. A multitude of Processing Elements (PEs) execute multiply-accumulations between ifmaps and filters and move data through custom interconnections. In addition, each PE can hold one or more types of data stationary to further reduce the energy consumption. For example, weight-stationary SAs move ifmaps and partial sums (psums) in orthogonal directions, while weights are stored and kept fixed at the PE level\cite{Wu_22}. When SAs process convolutions, PEs must consider the raster scan order of the ifmap activations. Indeed, 2D convolutions move $K \times K$ kernels over the ifmaps, from left to right and from top to bottom. Each movement configures a sliding window of activations, in turn belonging to consecutive ifmap rows. Adjacent windows share activations, thus data redundancy must be managed carefully. 

\begin{figure}
\includegraphics[width=\linewidth]{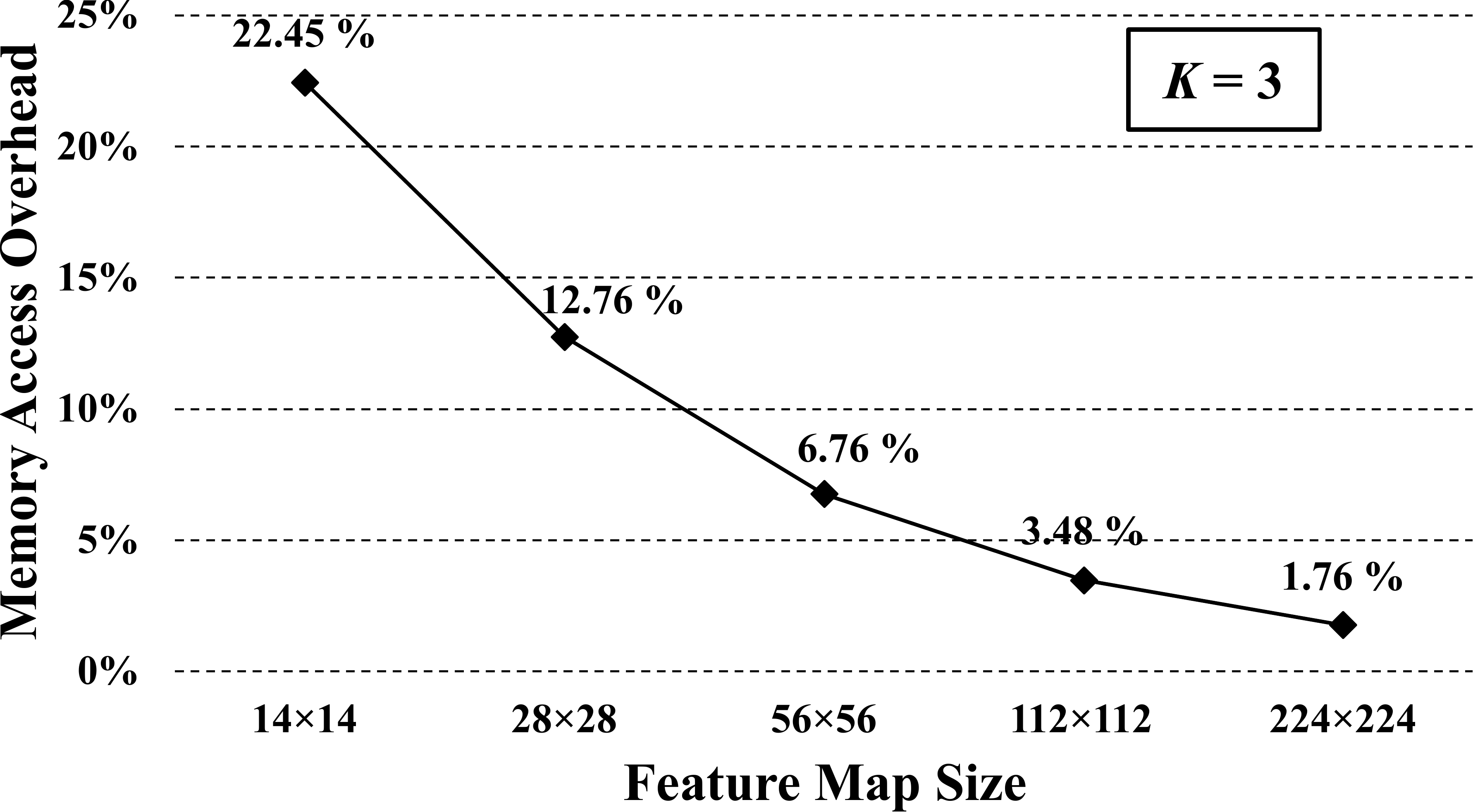}
\centering
\caption{Memory Access Overhead in TrIM\cite{Sestito_24_1}. The values relate to the processing of a single ifmap, with sizes as reported in the horizontal axis. The case of a $3 \times 3$ kernel is considered. Numbers are retrieved using the analytical model reported in \cite{Sestito_24_1}.
}
\label{MemAccOv_TrIM}
\end{figure}
SAs for CNNs can be classified into two groups: SAs using General Matrix Multiplication (GeMM-based SAs) and SAs directly dealing with the convolution workflow (Conv-based SAs). GeMM-based SAs\cite{Jouppi_17, Fornt_23, Wu_24, Feng_24} convert convolutions into matrix-multiplications and ifmaps need to be reorganized at the memory level, introducing data redundancy. Thus, GeMM-based SAs require larger memories and more memory accesses. Conv-based SAs\cite{Chen_17,Zhang_24,Sestito_24_1,Sestito_24_2} exploit custom dataflows to avoid redundancy at the memory level. In some architectures\cite{Chen_17,Zhang_24}, sliding windows are split in rows and reused at the PE level through scratch-pads (row-stationary dataflow\cite{Chen_16}). Despite this approach reduces the number of memory accesses, the continuous activity of scratch-pads degrades power, energy and area. To address this issue, authors in\cite{Sestito_24_1,Sestito_24_2} have recently proposed a local triangular dataflow, named TrIM, to maximize ifmap utilization without using scratch-pads at the PE level. After retrieving data vertically from the memory, activations are locally moved through horizontal and diagonal links using $K \times K$ PEs and shift register buffers to assist the raster scan order dictated by the convolution workflow. The dataflow has been validated in a hardware architecture where multiple TrIM slices are accommodated in cores to accelerate CNN workloads. However, the proposed architecture suffers from memory access overhead, since end-of-row ifmap activations must be read more than once from the memory. Figure~\ref{MemAccOv_TrIM} illustrates this overhead, considering convolutions on ifmaps having different sizes and $3 \times 3$ kernels. Notably, the overhead mainly affects small ifmaps, which are common in deep CNNs such as VGG-16\cite{Simonyan_15} and AlexNet\cite{Krizhevsky_12}. As a further drawback, each TrIM slice\cite{Sestito_24_2} requires independent shift register buffers due to the specific ifmap orientation. No buffer sharing is implemented at the core level, thus degrading area, power and energy.

To address the referred problems, we propose an upgraded version, named 3D-TrIM. The architecture proposed here introduces few shadow registers at the buffer level to reuse end-of-row activations without further access to the external memory. Furthermore, 3D-TrIM uses a different architectural orientation to reduce the number of shift register buffers: each core processes one ifmap through different filters, thus needing a single local buffer to store activations for reuse. Finally, the orientation change affects the additional adder trees for spatial accumulations, which are now shared among the different cores, resulting in a 3D architecture. 

The specific contributions of this work are:
\begin{itemize}
    \item A 3D systolic array architecture for CNNs to maximize ifmap utilization. The proposed architecture exploits local buffer sharing for energy and area efficiency. 
    \item The use of efficient shadow registers to nullify the memory access overhead at the ifmap level. These registers are part of the local buffer resources.
    \item An architecture of 576 PEs is implemented on commercial 22 nm technology. The peak throughput is 1.15 TOPS, with a 0.26 mm$^2$ area occupation, and 0.25 W power dissipation. 3D-TrIM outperforms TrIM in terms of operations per memory access up to $3.37\times$ on VGG-16 and AlexNet CNNs.
\end{itemize}

The rest of the paper is organized as follows: Section II introduces the architecture of 3D-TrIM. Implementation and characterization results are presented in Section III. Finally, Section IV concludes the paper.

\section{The 3D-TrIM Architecture}

\begin{figure}
\includegraphics[width=\linewidth]{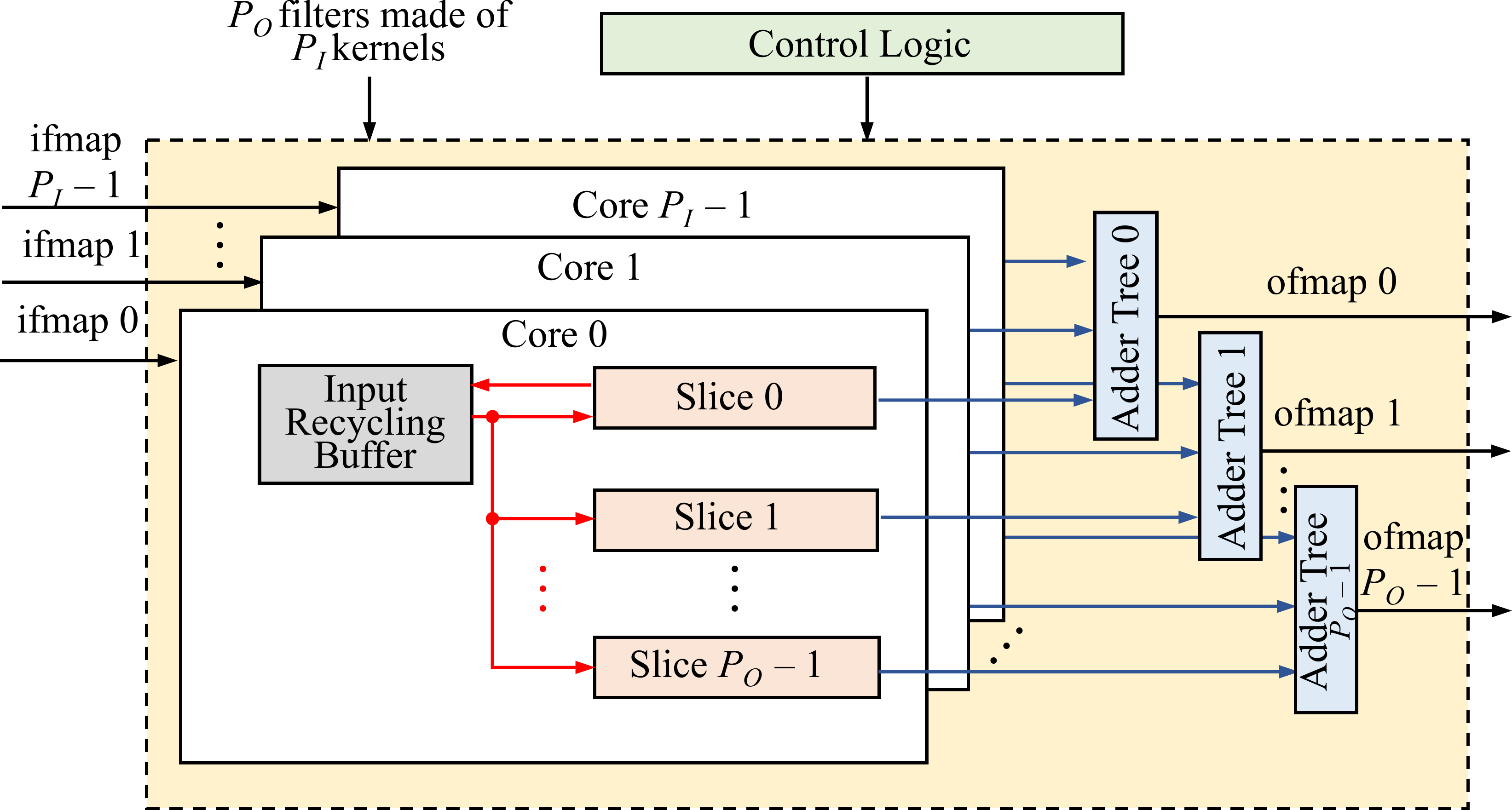}
\centering
\caption{Top-level architecture of 3D-TrIM. $P_I$ cores execute convolutions between $P_I$ ifmaps and $P_I \times P_O$ kernels. Each core hosts $P_O$ slices, operating on the same ifmap. To maximize ifmap utilization on-chip, an Input Recycling Buffer is accommodated at the core level and shared among the slices. $P_O$ adder trees accumulate psums to finalize the convolution. A control logic supervises the functionality of the entire architecture over time.
}
\label{3D_SA_TOP}
\end{figure}
The top-level architecture of 3D-TrIM is shown in Fig.~\ref{3D_SA_TOP}. $P_I$ parallel cores perform $P_I \times P_O$ convolutions between $P_I$ ifmaps and $P_I \times P_O$ kernels, with $P_I$ being the number of ifmaps processed in parallel, and $P_O$ being the number of 3D filters processed in parallel (or, equivalently, the number of ofmaps generated in parallel). Each core hosts $P_O$ slices coping with $K \times K$ convolutions. To maximize ifmap utilization on-chip, each core includes an Input Recycling Buffer (IRB). Slice 0 supplies the IRB with ifmap activations. When required, those activations are broadcast to the other slices of the core. In this way, ifmap activations are read once from the memory and then reused locally as long as required. $P_O$ adder trees are placed straight after the cores in order to accumulate the temporary psums. A control logic provides configuration and enable signals to buffers and processing units.

\subsection{The Slice}

\begin{figure}
\includegraphics[width=\linewidth]{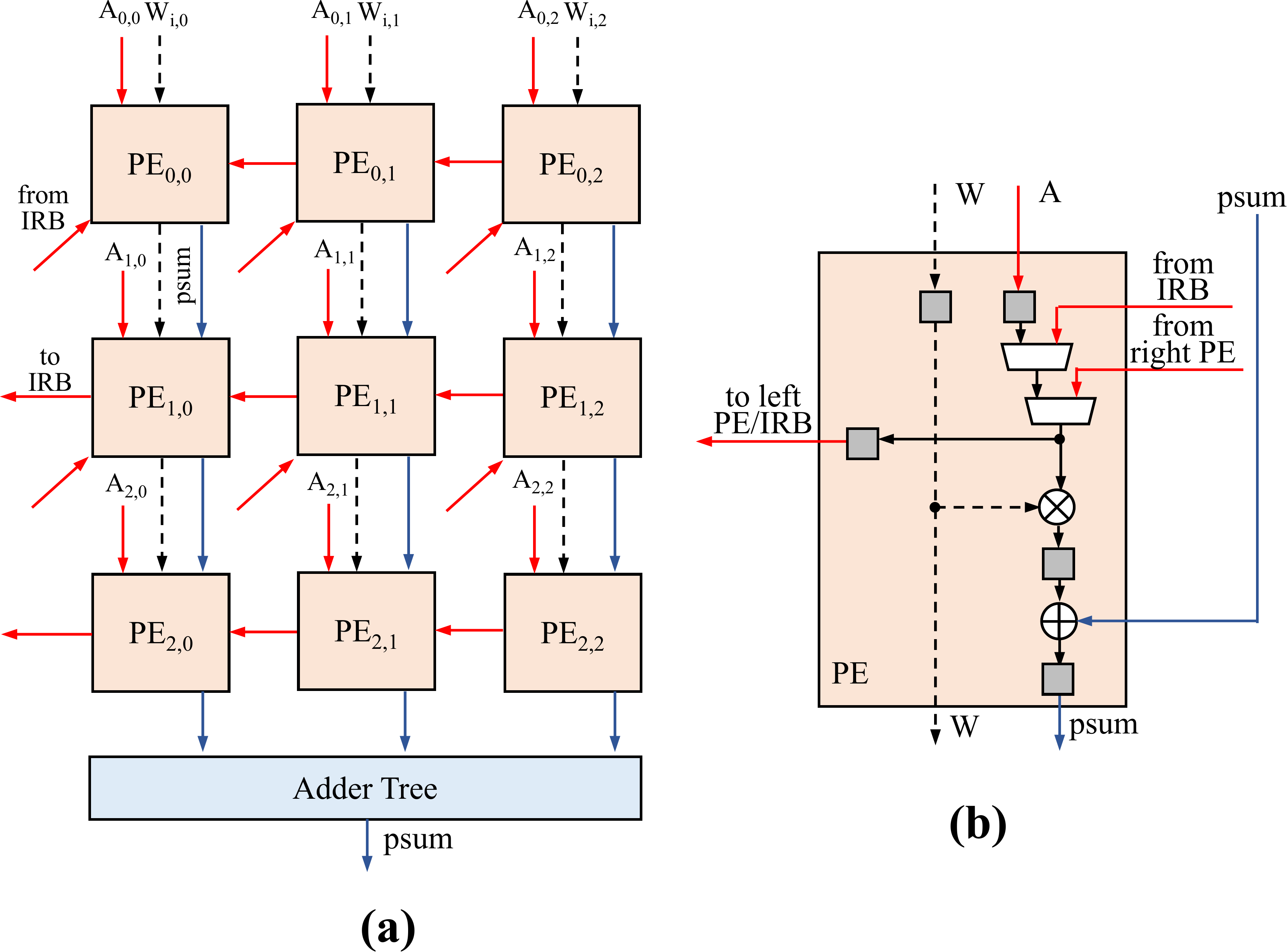}
\centering
\caption{The slice when $K=3$. (a) It consists of $3 \times 3$ Processing Elements (PE), interconnected with each other in vertical and horizontal directions. In addition, the slice interacts with the Input Recycling Buffer to provide and read back activations to be reused. The read back activity is finalized by diagonal connections. An adder tree eventually accumulates psums coming from bottom PEs. Red arrows relate to activations; black dashed arrows relate to weights; blue arrows relate to psums.
(b) Each PE stores the current activation, weight, psum in registers. Two multiplexers select the direction of the activation to be reused (vertically, horizontally, diagonally). Finally, a pipelined multiply-accumulation unit executes the current computation.
}
\label{SA_PE}
\end{figure}
Each slice consists of $K \times K$ PEs and an adder tree. Figure~\ref{SA_PE} shows the example of a slice when $K=3$. The generic PE, other than hosting a multiply-accumulation unit, contains registers to store the external activation, weight, product, psum, and the activation to be moved to the left-adjacent PE/IRB in the next cycle. Two multiplexers select the direction of the current activation, which can be either supplied by the memory, or by the PE placed on the right side, or reused from the IRB.
The dataflow is the following:
\begin{itemize}
    \item Weights are fetched from the memory, loaded vertically into the PEs placed at the top boundary of the array, and then moved from top to bottom. At the completion, these weights are kept stationary at the PE level.
    \item Activations are fetched from the memory and supplied vertically to each PE. Then, these activations are moved from right to left. Once activations reach the leftmost column of the array, these can be either forwarded to the IRB (in the case of Slice 0) or discarded. Finally, activations coming from the IRB are forwarded to the PEs for further reuse. 
    \item Psums move from top to bottom over consecutive cycles. Bottom psums are eventually summed up through an adder tree.
\end{itemize}

\subsection{Input Recycling Buffer}

\begin{figure}
\includegraphics[width=\linewidth]{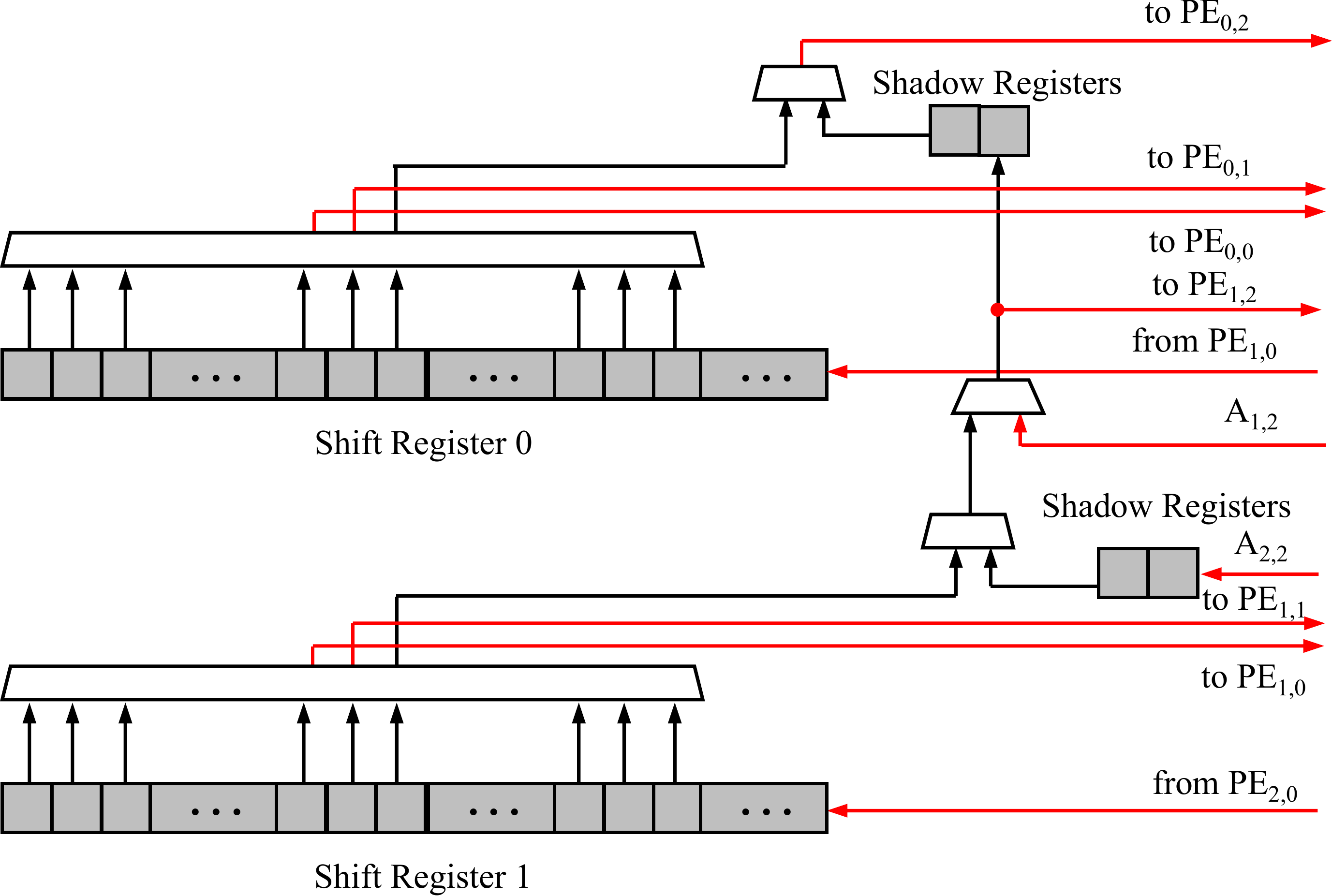}
\centering
\caption{The Input Recycling Buffer when $K=3$. It consists of two reconfigurable shift registers, shadow registers and multiplexers. Shift registers read activations from Slice 0 (in each core). After some cycles, these shift registers provide activations back to PEs for reuse. Shadow registers manage end-of-row ifmap activations. Multiplexers select whether the current activations must be provided by shift-registers or by shadow registers. 
}
\label{IRB}
\end{figure}
The IRB assists the slices to provide the set of activations that need to be reused over different cycles. It consists of (i) $K-1$ shift registers (each accommodating $W_I-K-1$ registers, where $W_I$ is the ifmap width); (ii) $(K-1) \times (K-1)$ shadow registers; (iii) multiplexers. Figure~\ref{IRB} depicts the architecture of the IRB when $K=3$. For each sliding window covering three ifmap rows, Shift Register 0 keeps track of the second row under processing, while Shift Register 1 manages the third row. Both shift registers do not manage the last two activations of each ifmap row, for which shadow registers are responsible.
With the aim to support different ifmap sizes, shift registers have been made reconfigurable. To achieve this, internal stages of the shift registers are connected to a multiplexing logic that routes the correct set of activations to the slices. The other multiplexers reported in Fig.~\ref{IRB} control whether activations are provided by shift registers or by shadow registers.

\begin{figure*}
\includegraphics[width=0.9\textwidth]{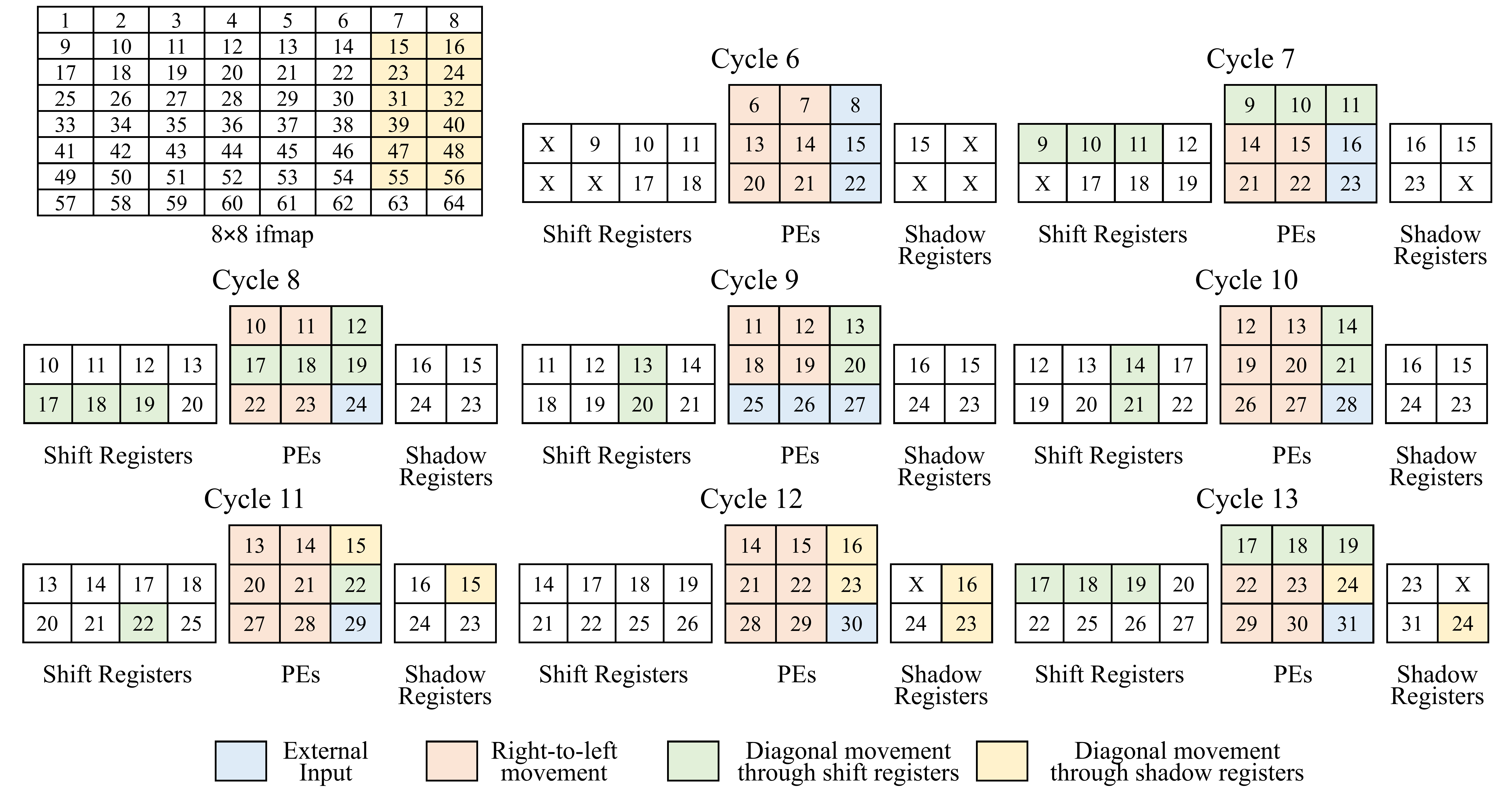}
\centering
\caption{Example of dataflow. An $8 \times 8$ ifmap is considered. The computational cycles from 6 to 13 are visualized in detail. In the ifmap, the area in yellow indicates the portion managed by the shadow registers. For each cycle, the activations processed by the PEs, shift registers and shadow registers are reported. Activations in blue are read from the memory. Activations in orange are shifted through right-to-left movements. Activations in green relate to diagonal movements through shift registers. Activations in yellow relate to diagonal movements using shadow registers. Xs refer to don't care cases.
}
\label{IRB_ex}
\end{figure*}
To facilitate the understanding of the IRB, let consider the case reported in Fig.~\ref{IRB_ex}. The example  illustrates the convolution between an $8 \times 8$ ifmap and a $3 \times 3$ kernel, with a focus on the activity between cycle 6 and cycle 13. For each cycle, the activations being processed by the PEs are reported, as well as the activations stored in the shift registers and in the shadow registers. Throughout the cycles, activations can be supplied to PEs externally (activations in blue), horizontally (activations in orange), diagonally. The IRB assists the systolic array with diagonal movements: shift registers (activations in green) and shadow registers (activations in yellow) can feed the PEs. For each ifmap row to be reused, shift registers host the majority of activations, except for the end-of-row ones. For example, at cycle 7, activations 9, 10, 11 are reused by PE$_{0,0}$, PE$_{0,1}$, PE$_{0,2}$, respectively, through shift registers. Conversely, shadow registers store end-of-row activations and constitute the key novelty of the IRB proposed in 3D-TrIM. In detail, between cycles 6 to 8, the activations 15, 16, 23, 24 are stored in shadow registers for reuse in the next end-of-row sliding windows. Such activations are restored in cycles 11-13: activations 15, 16 are supplied to PE$_{0,2}$, while activations 23, 24 are supplied to PE$_{1,2}$. Activations 23, 24 are also shifted between shadow registers to allow further reuse in the next end-of-row sliding windows (not shown in Fig.~\ref{IRB_ex}).

\subsection{Adder Trees and Control Logic}

With reference to the CNN workload, each core processes one ifmap with a set of kernels belonging to different filters. In order to generate an ofmap, the psums generated by the different cores need to be spatially accumulated. For this reason, $P_O$ adder trees are interfaced with the cores, and each adder tree sums up $P_I$ operands. For example, based on Fig.~\ref{3D_SA_TOP}, Adder Tree 0 manages psums generated by each Slice 0 in the different cores. Similarly, Adder Tree 1 is responsible for psums provided by each Slice 1. And so on until Adder Tree $P_O-1$.

The control logic orchestrates the operations of 3D-TrIM over time. At the slice level, the control logic manages the selection signals of the multiplexers to load the activations from the correct direction. The proper subset of activations from shift registers and the activity of shadow registers are handled at the IRB level.

\section{Implementation and Characterization}

3D-TrIM is implemented on commercial 22 nm technology. The architecture is designed in Verilog, synthesized by Cadence Genus, and placed and routed by Cadence Innovus. The parallelism parameters are set to $P_I=8$ and $P_O=8$ (576 PEs), while the clock frequency is set to 1 GHz. According to these constraints, the design achieves a peak throughput of 1.15 Tera Operations per Second (TOPS), occupying an area of 0.26 mm$^2$ and dissipating 0.25 W. 

\begin{figure*}
\includegraphics[width=0.9\textwidth, height=5.5cm]{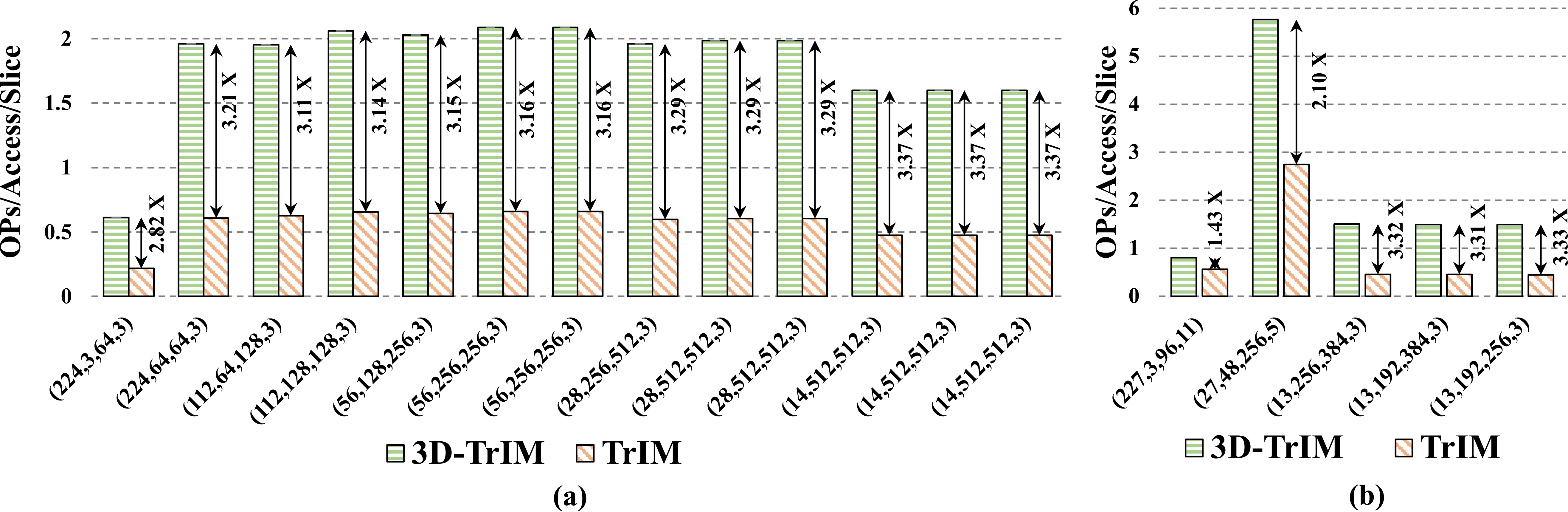}
\centering
\caption{3D-TrIM vs TrIM\cite{Sestito_24_2}: Comparisons on the (a) VGG-16 and (b) AlexNet CNNs. For each convolution layer, the horizontal label indicates the ifmap size $I$, the number of input channels $C$, the number of filters $F$, and the kernel size $K$ in the form $(I,C,F,K)$. The values relate to the number of operations per memory access per slice (OPs/Access/Slice) experienced by each layer. The green bars refer to 3D-TrIM, while the orange bars refer to TrIM.
}
\label{OPs_MemAcc}
\end{figure*}
Figure~\ref{OPs_MemAcc} compares 3D-TrIM and TrIM\cite{Sestito_24_2} in terms of operations per memory access per slice (OPs/Access/Slice). The normalization is done considering that 3D-TrIM uses $8 \times 8$ slices, while TrIM uses $7 \times 24$ slices. The VGG-16\cite{Simonyan_15} and AlexNet\cite{Krizhevsky_12} CNNs are referenced as case studies. The feature extraction section of VGG-16 consists of 13 convolution layers with ifmap sizes ranging from $224\times224$ to $14\times 14$ and kernel size fixed to $3\times3$. AlexNet features 5 convolution layers with ifmap sizes ranging from $227\times227$ to $13\times13$, and kernel sizes ranging from $11\times11$ to $3\times3$. It is worth underlining that 3D-TrIM can support kernel sizes larger than $3 \times 3$ through kernel tiling. For example, a $5 \times 5$ kernel can be split into four $3 \times 3$ sub-kernels. The different sub-kernels are assigned to as many cores. The adder trees are later responsible for accumulating the psums from each sub-kernel. According to Fig.~\ref{OPs_MemAcc}, 3D-TrIM outperforms TrIM\cite{Sestito_24_2} given the effectiveness of shadow registers to mitigate the memory access overhead. The improvement is in the range $2.82\times$--$3.37\times$ for VGG-16, and in the range $1.43\times$--$3.33\times$ for AlexNet. Furthermore, it should be noted that 3D-TrIM executes the same set of operations using $2.6\times$ fewer slices than TrIM\cite{Sestito_24_2}, in addition to saving IRBs due to the sharing approach at the core level.

\begin{table}[t]
  \centering
   \setlength{\tabcolsep}{5pt} 
   \renewcommand{\arraystretch}{0.7} 
  \caption{State-of-The-Art Architectures}
  \begin{tabular}{c c c c c}
  \toprule
    \ & \cite{Jouppi_21} & \cite{Chen_17} & \cite{Feng_24} & 3D-TrIM \\
    \midrule
     \\ Number of PEs & 65536 & 168 & 256 & 576 \\
     \\ Technology [nm] & 7 & 65 & 7 & 22 \\
     \\ Frequency [GHz] & 1.05 & 0.2 & 2 & 1 \\
     \\ Peak Throughput [TOPS] & 138 & 0.07 & 1.02 & 1.15 \\
     \\ Norm. Peak Throughput$^{\mathrm{*}}$ [TOPS] & 117.55 & 0.11 & 0.87 & 1.15 \\
     \\ Area [mm$^2$] & $<$400 & 12.25 & 3.81 & 0.26 \\
     \\ Norm. Area$^{\mathrm{*}}$ [mm$^2$] & $<$8000 & 1.32 & 76.12 & 0.26 \\
     \\ Power [W] & 175 & 0.24 & 5.12 & 0.25 \\
     \\ Norm. Power$^{\mathrm{*}}$ [W] & 399.54 & 0.11 & 11.70 & 0.25 \\
     \\ Norm. Energy Eff.$^{\mathrm{*}}$ [TOPS/W] & 0.29 & 0.96 & 0.07 & 4.54 \\
     \\ Norm. Area Eff.$^{\mathrm{*}}$ [TOPS/mm$^2$] & $>$0.01 & 0.08 & 0.01 & 4.47 \\
\bottomrule 
    \multicolumn{5}{l}{$^{\mathrm{*}}$Normalized to 22 nm technology using the DeepScaleTool\cite{Sarangi_1,Sarangi_2}.} \\
  \end{tabular}
  \label{SOTA}
\end{table}
Table~\ref{SOTA} collects architectural information for both 3D-TrIM and previous works\cite{Jouppi_21,Chen_17,Feng_24} implemented in ASIC. Number of PEs, peak throughput, area, and power are reported. Energy and area efficiency are also reported, by normalizing data to 22 nm using the DeepScaleTool\cite{Sarangi_1,Sarangi_2}. 

The architecture in \cite{Jouppi_21} refers to the Google Tensor Processing Unit (TPU) v4i. This version includes four Matrix Multiply Units (MXUs), each featuring a systolic array of $128\times128$ PEs. The architecture, running at a frequency of 1050 MHz, achieves a peak throughput of 138 TOPS. However, Google TPU v4i dissipates 175 W, thus degrading the energy efficiency metric.

Eyeriss\cite{Chen_17} is a spatial array of $12\times14$ PEs running at 200 MHz. It is based on the row-stationary dataflow, where ifmap and weights are reused at the PE level through scratch-pads. The same set of data is broadcast between PEs for parallel computations. The intense activity of scratch-pads degrades power and energy, thus limiting the efficiency of the architecture. On the contrary, 3D-TrIM uses straightforward PEs where weights are stationary, while ifmap activations move at the SA level through the triangularity of the dataflow.

The SA presented in \cite{Feng_24} consists of $16 \times 16$ PEs following the weight-stationary dataflow and running at a frequency of 2 GHz. This architecture includes a logic to cope with multi-precision computations to enhance the PE utilization, other than requiring synchronization FIFOs as dictated by the dataflow. The above requirements result in higher energy and larger area.

\section{Conclusion}

Next-generation AI hardware aims at mitigating the Von Neumann bottleneck, thus reducing the cost to move data from memory to core and maximizing the energy efficiency. Systolic arrays are promising architectures in this direction, because they provide interconnected processing elements for local data reuse. Convolutional Neural Networks can benefit from systolic arrays to manage multi-dimensional ifmaps and filters properly.

This paper presents 3D-TrIM: a systolic array for energy- and area-efficient convolutions. Several 2D slices interact with few Input Recycling Buffers to eliminate the memory access overhead at the ifmap level and to enable buffer sharing. Each buffer accommodates shift registers and few extra shadow registers. An architecture of 576 PEs is implemented on commercial 22 nm technology and guarantees an energy efficiency of 4.54 TOPS/W and an area efficiency of 4.47 TOPS/mm$^2$. In addition, 3D-TrIM outperforms TrIM by up to $3.37\times$ in terms of operations per memory access on VGG-16 and AlexNet CNNs.

\bibliographystyle{IEEEtran}
\bibliography{IEEEabrv,bib}

\end{document}